\documentclass[figures]{epl}
\usepackage{amsmath}
\usepackage{amssymb}

\title{Propagation of small perturbations in synchronized
oscillator networks}
\shorttitle{Perturbation propagation in oscillator networks}
\author{Dami\'an H. Zanette}
\shortauthor{D. H. Zanette} \institute{Consejo Nacional de
Investigaciones Cient\'{\i}ficas y T\'ecnicas, Centro At\'omico
Bariloche and Instituto Balseiro, 8400 Bariloche, R\'{\i}o Negro,
Argentina }


\begin{document}

\maketitle

\begin{abstract}
We study the propagation of a harmonic perturbation of small
amplitude on a network of coupled identical phase oscillators
prepared in a state of full synchronization. The perturbation is
externally applied to a single oscillator, and is transmitted to
the other oscillators through coupling. Numerical results and an
approximate analytical treatment, valid for random and ordered
networks, show that the response of each oscillator is a rather
well-defined function of its distance from the oscillator where
the external perturbation is applied. For small distances, the
system behaves as a dissipative linear medium: the perturbation
amplitude decreases exponentially with the distance, while
propagating at constant speed. We suggest that the pattern of
interactions may be deduced from measurements of the response of
individual oscillators to perturbations applied at different
nodes of the network.
\end{abstract}

Synchronization phenomena have recently attracted a great deal of
attention \cite{bocca1}. The emergence of synchronous behaviour in
large ensembles of interacting dynamical elements is a
paradigmatic form of collective self-organization, typical of a
vast class of natural systems ranging from complex chemical
reactions to large-scale biological processes \cite{Win,Mikh}.
Many of these phenomena are successfully reproduced by relatively
simple mathematical models. The most basic form of synchronized
dynamics, which can be achieved in an ensemble of identical
periodic oscillators subject to attractive coupling, is full
synchronization \cite{nos}. In such state, the individual motions
of all oscillators coincide.

A network of fully-synchronized coupled oscillators may be viewed
as an active medium in a highly coherent collective state. A
relevant question regarding the dynamics of this system is how
the medium responds to an external perturbation which affects its
coherence. Due to the coupling between oscillators, the
perturbation should spread through the medium. Moreover, if
dissipative mechanisms are acting and the state of full
synchronization is stable, the perturbation should die out as the
distance from the point where it is applied becomes larger. The
properties of this propagation phenomenon provide a dynamical
characterization of the synchronized medium, much like the
propagation of a defect in an extended system. This is the
problem that we address in this Letter.

Kuramoto's model of coupled phase oscillators \cite{Kura} provides
the basis  for our model. The state of each oscillator is
described by a phase variable $\phi_i \in [0,2\pi)$ which, in the
absence of coupling, rotates at constant frequency, $\dot
\phi_i=\omega_i$. A network of coupled oscillators is governed by
the equations
\begin{equation} \label{osc1}
\dot \phi_i = \omega_i + k \sum_{j=1}^N J_{ij} \sin
(\phi_j-\phi_i),
\end{equation}
where $k$ is the coupling strength. Here, $J_{ij}=1$ if
oscillator $i$ is coupled to oscillator $j$, and $J_{ij}=0$
otherwise. The connection (or adjacency) matrix ${\cal J}=\{
J_{ij} \}$ is not necessarily symmetric. The underlying connection
network is therefore a directed graph, where the links pointing
to the node occupied by a given oscillator $i$ start at those
oscillators which enter the equation of motion for $\phi_i$.

If the oscillators are identical, $\omega_i=\omega_j$ for all $i$,
$j$. Without generality loss, their natural frequencies are
chosen to be $\omega_i=0$, and the coupling strength is fixed at
$k=1$. To represent the external perturbation, one of the
oscillators --say, $i=1$-- is also coupled to an oscillating
force of strength $\epsilon$ and frequency $\Omega$. Starting
from eq.~(\ref{osc1}), the equation of motion for $\phi_i$ can be
written as
\begin{equation} \label{osc2}
\dot \phi_i =  \sum_{j=1}^N J_{ij} \sin (\phi_j-\phi_i) +
\epsilon \delta_{i1} \sin (\Omega t-\phi_1),
\end{equation}
where $\delta_{ij}$ is the Kronecker symbol. In the absence of
external forcing ($\epsilon=0$) the ensemble admits a fully
synchronized state where $\phi_i = \phi^*$ for all $i$, with
constant $\phi^*$. The stability of this state can be
analytically proven for some specific forms of the connection
matrix $\cal J$. In particular, full synchronization is stable in
the case where the number of non-vanishing elements in all rows of
$\cal J$ is the same, i.e. when the number $z_i=\sum_j J_{ij}$ of
connections ending at each oscillator is the same, $z_i=z$, for
all $i$ \cite{str}. In this case, the network is a regular
directed graph \cite{Gross}. The following study is mainly
focused on this kind of network.

A small perturbation of the fully synchronized state produces a
deviation of the same order $\epsilon$ as the perturbation. For
$\epsilon \to 0$, the solution to eq.~(\ref{osc2}) can be found by
writing $\phi_i = \phi^*+ \epsilon \psi_i$, and expanding up to
the first order in $\epsilon$, which yields
\begin{equation} \label{osc3}
\dot \psi_i =  \sum_{j=1}^N J_{ij} (\psi_j-\psi_i)+ \delta_{i1}
\exp({\rm i} \Omega t).
\end{equation}
In the last term,  we have replaced $\sin (\Omega t)$ by
$\exp({\rm i} \Omega t)$, for convenience in the mathematical
treatment. After transients have elapsed, the solution to eq.
(\ref{osc3}) has the form $\psi_i(t) = A_i \exp({\rm i} \Omega
t)$. The complex amplitude $A_i$ is obtained from the linear
system
\begin{equation} \label{calM}
{\cal M} {\bf A} = {\bf e}_1,
\end{equation}
where ${\bf A}=(A_1,A_2,\dots, A_N)$, ${\bf e}_1=(1,0, \dots,0)$,
and ${\cal M}=(z+{\rm i}\Omega){\cal I}-{\cal J}$, with $\cal I$
the $N\times N$ identity matrix. The amplitudes are, thus, ${\bf
A}={\cal M}^{-1} {\bf e}_1$. For a given connection matrix $\cal
J$, they can be found by numerically inverting $\cal M$.

\begin{figure}
\twoimages[scale=0.3]{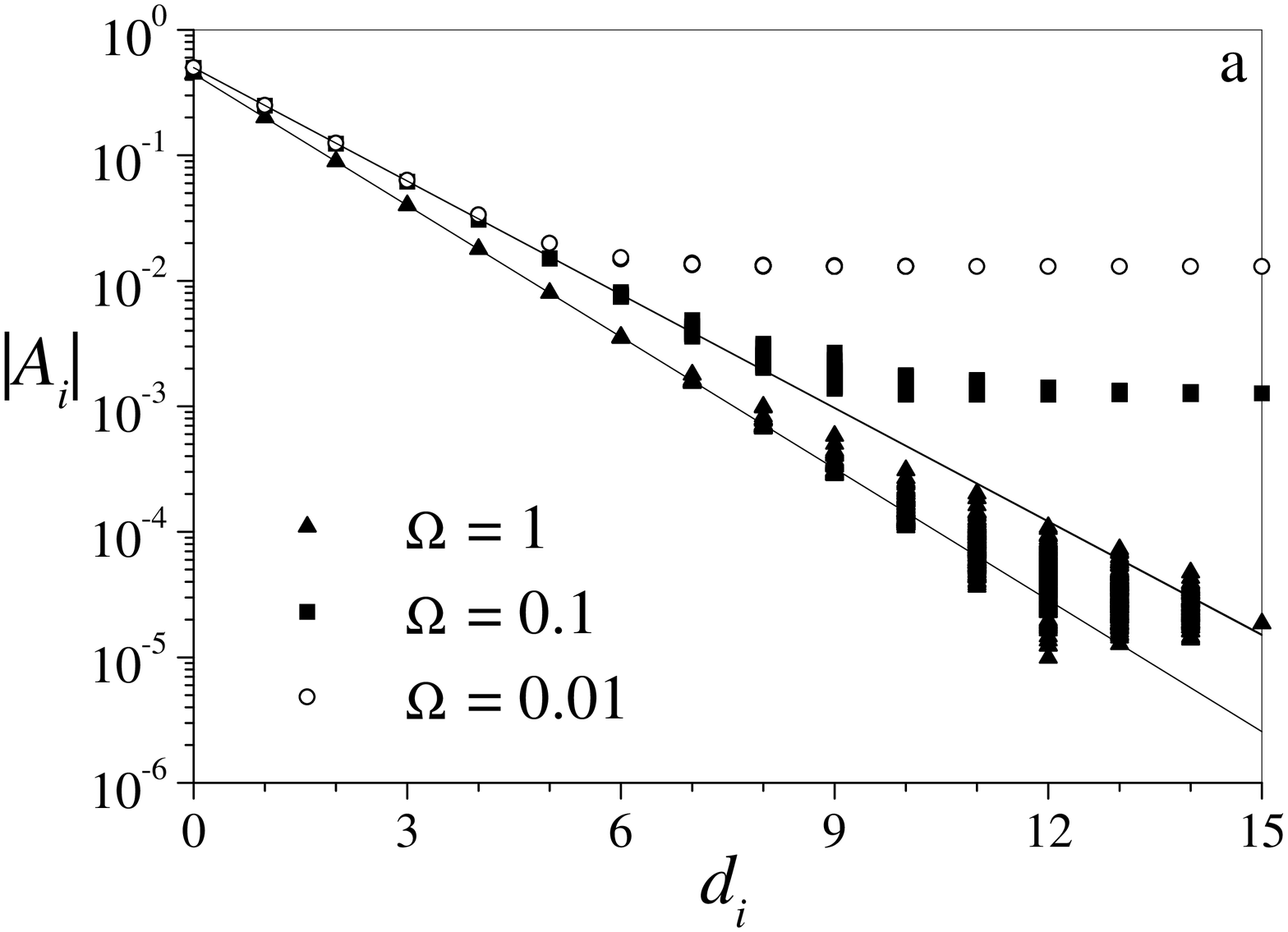}{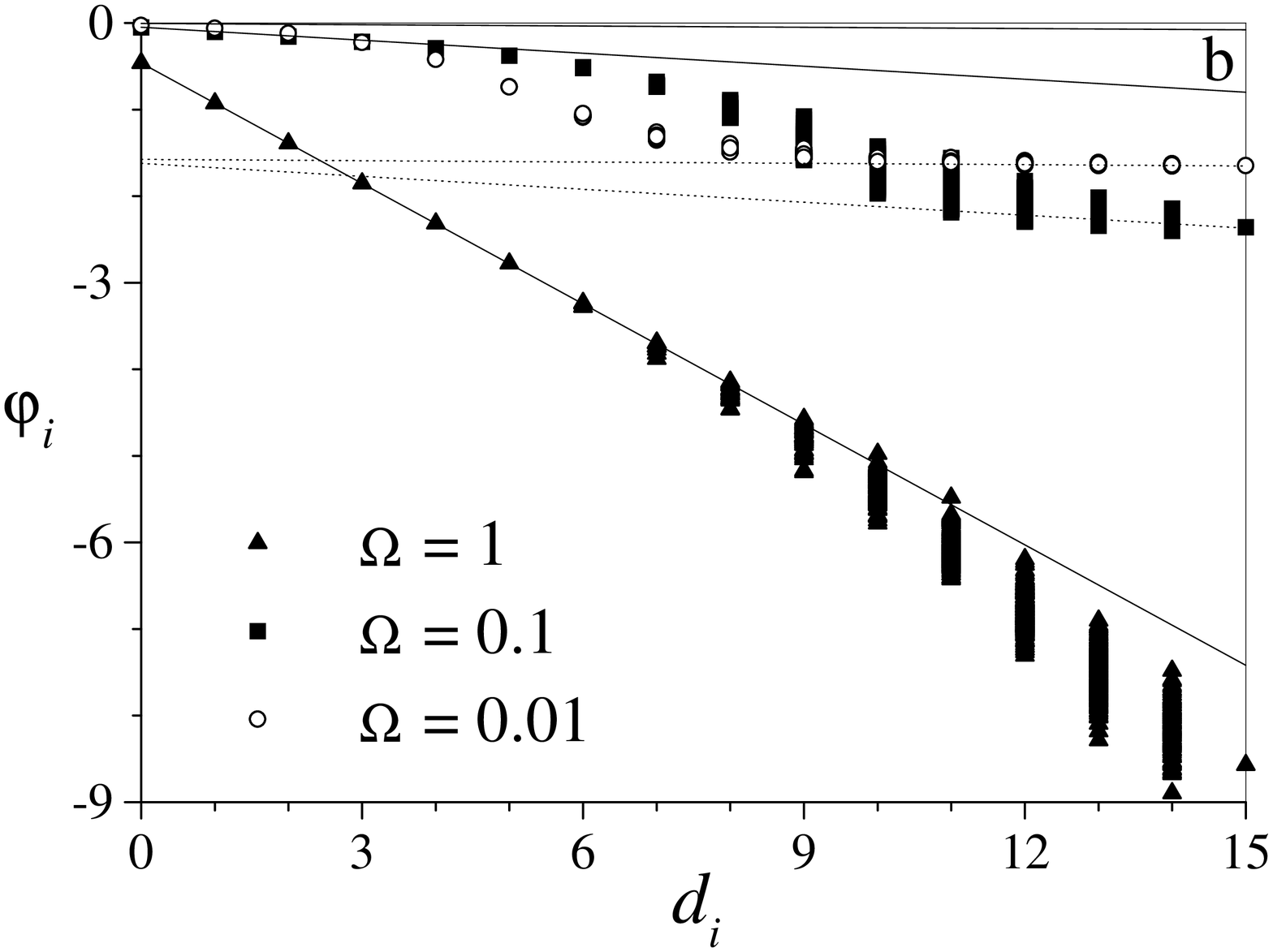} \caption{Moduli (a) and
phases (b) of the individual amplitudes $A_i$ in a random network
($z=2$) of $10^3$ fully synchronized oscillators, as functions of
the distance to the perturbed oscillator. Each dot corresponds to
the amplitude of a single oscillator.  The three data sets have
been obtained for the same realization of the network but with
different perturbation frequencies $\Omega$. Full and dotted
lines are analytical predictions for small and large distances,
respectively.} \label{fig1}
\end{figure}

Figure \ref{fig1} shows typical results for the moduli $|A_i|$
and the phases $\varphi_i$ of the amplitudes, $A_i\equiv |A_i|
\exp({\rm i} \varphi_i)$, in a $10^3$-oscillator random network.
Each oscillator is coupled to $z=2$ neighbours, which are chosen
at random from the whole ensemble (avoiding self and multiple
connections). The three data sets of the figure show results for
the same random network and different perturbation frequencies
$\Omega$. Each dot represents the modulus or phase of a single
oscillator $i$ as a function of its distance $d_i$ to oscillator
$1$, where the external perturbation is applied. This distance is
defined as the number of connections along the shortest
(directed) path starting at oscillator $1$ and ending at $i$. For
this particular network, $d_i$ varies between $0$ (for $i=1$) and
$15$.

The numerical results of fig.~\ref{fig1} have been obtained by
inversion of the matrix $\cal M$ in eq.~(\ref{calM}). We have
verified that, as expected, these results are reproduced by
numerical integration of the equations of motion (\ref{osc2}), for
sufficiently small values of the perturbation amplitude
($\epsilon \lesssim 10^{-4}$).

It turns out that the modulus $|A_i|$ has a rather well-defined
dependence on the distance $d_i$. For small distances, it decays
exponentially, and its value is essentially the same for all the
oscillators at a given distance. As $d_i$ grows, however, the values
of $|A_i|$ become more dispersed as a function of $d_i$, especially
for large perturbation frequencies. Moreover, the exponential decay
breaks down, and the variation of $|A_i|$ with the distance becomes
much slower. The values of $|A_i|$ for large distances depend on the
frequency $\Omega$. Numerical results for different realizations of
the random network show that these values can strongly differ
between realizations, and seem to be determined by the maximum value
of $d_i$ in a given network.

As shown in fig.~\ref{fig1}b, the phase $\varphi_i$ is negative
for all distances and its absolute value increases with $d_i$.
This implies that the response of oscillators to the perturbation
is increasingly retarded as the distance grows. For small
distances we find a linear regime, with a constant phase shift
$\Delta \varphi$ between oscillators whose distances differ by
one. Consequently, the perturbation propagates at constant speed
$v=|\Omega/\Delta \varphi|$. For larger distances, before the
regime of exponential decay in $|A_i|$ breaks down
(cf.~figs.~\ref{fig1}a and b), this linear dependence is lost.
After an intermediate zone of faster variation, where the
dispersion of the phase as a function of the distance is larger,
the dependence of $\varphi_i$ on $d_i$ becomes much less
pronounced. The linear regime is particularly well defined for
large perturbation frequencies, while the large-distance
behaviour is clearly seen for small $\Omega$.

The dependence of $A_i$ on  $d_i$ can be understood by relating
the connection matrix $\cal J$ with the distribution of distances
in the network. The starting point is the solution to eq.
(\ref{calM}), ${\bf A}={\cal M}^{-1} {\bf e}_1$. Since the
eigenvalues of the connection matrix $\cal J$ are all less than
or equal to $z$ in modulus \cite{str}, the matrix ${\cal
M}^{-1}=[(z+{\rm i}\Omega){\cal I}-{\cal J}]^{-1}$ can be
expanded as a series of powers of $\cal J$ (for $\Omega\neq 0$).
The amplitude of the deviation from the fully synchronized state
for the $i$-th oscillator turns out to be
\begin{equation} \label{Ai}
A_i = \sum_{m=0}^\infty  (z+{\rm i}\Omega)^{-m-1} J^{(m)}_{i1},
\end{equation}
where $J^{(m)}_{ij}$ is an element of the $m$-th power of the
connection matrix: ${\cal J}^m = \{ J^{(m)}_{ij} \}$. The matrix
${\cal J}^m$ bears explicit information about the metric
structure of the network. Specifically, the element $J^{(m)}_{ij}$
equals the number of directed paths of length $m$ starting at
node $j$ and ending at node $i$ \cite{Gross}. Taking $j=1$, i.e.
the node at which the external perturbation is applied, we have
in particular that $J^{(m)}_{i1}=0$ for $m<d_i$ and
$J^{(m)}_{i1}\neq 0$ for $m=d_i$. If $m>d_i$, $J^{(m)}_{i1}$ is
zero if no path of length $m$ joins oscillators $1$ and $i$, and
positive otherwise.

For oscillator $i$, thus, the first contribution to the sum in eq.
(\ref{Ai}) is the term with $m=d_i$. If $z/N \ll 1$ and $d_i$ is
small, there is essentially only one path of length $d_i$ from $1$
to $i$. This is realized by noticing that the probability of
having more than one path of small length between two given
oscillators is, at most, of order $z/N$. Consequently, for the
majority of nodes at small distances from the perturbed
oscillator, we have $J^{(d_i)}_{i1}=1$. The total fraction
$n(d_i)$ of nodes with small $d_i$ is also small, $n(d_i)\approx
z^{d_i}/N$ and, therefore, the possibility that a node at
distance $d_i$ is also connected by a path of length slightly
larger than $d_i$ can be neglected. This implies that, for
oscillators at small distances $d_i$, the sum in eq. (\ref{Ai})
is dominated by the first nonzero term:
\begin{equation} \label{Ai1}
A_i \approx (z+{\rm i}\Omega)^{-d_i-1}
=(z^2+\Omega^2)^{-\frac{d_i+1}{2}} \exp \left[ -{\rm i} (d_i+1)
\tan^{-1}   \frac{\Omega}{z} \right].
\end{equation}

This dominance is enhanced as $|z+{\rm i}\Omega|$ becomes large,
since higher-order terms are weighted by increasing powers of this
number. In particular, the contribution of the first nonzero term
becomes increasingly important as the frequency $\Omega$ grows. In
the right-hand side of eq. (\ref{Ai1}), the exponential dependence of
$|A_i|$ and the linear dependence of $\varphi_i$ on $d_i$, with a
phase shift $\Delta \varphi = \tan^{-1} (\Omega/z)$, are apparent.
Full straight lines in fig. \ref{fig1} show the quantitative
agreement of this prediction with numerical results for small
distances. Note that, as expected, the exponential approximation
improves for larger $\Omega$.

\begin{figure}
\onefigure[width=10cm]{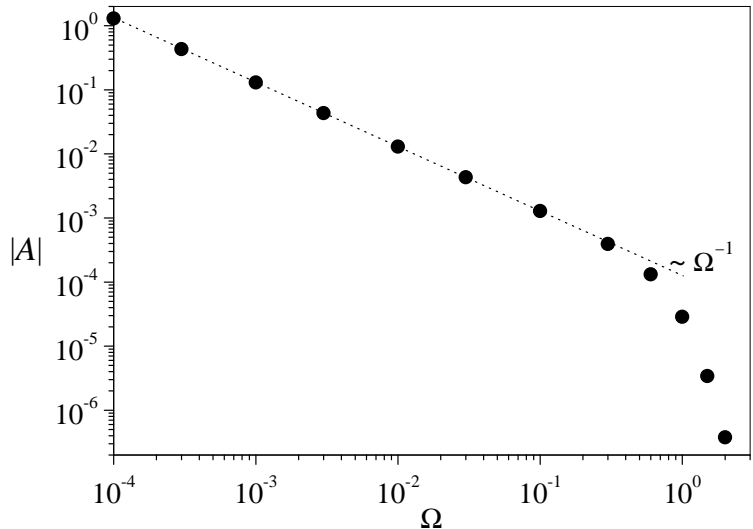} \caption{Amplitude modulus at
the maximal distance, $d_i=15$, for the $10^3$-oscillator random
network considered in fig. \ref{fig1}, as a function of the
perturbation frequency $\Omega$. The dotted line has slope $-1$.}
\label{fig2}
\end{figure}

For large distances, such that $z^{d_i} \sim N$, it is not any
longer possible to insure that the first contribution to $A_i$ is
given by only one path of length $d_i$, nor that higher-order
contributions are relatively negligible. In fact, one can argue
that the number of paths of length $m$ ending at a given node $i$
scales as $J^{(m)}_{i1} \sim z^m$ for large $m$. This can be
proven inductively, noting that this number is $z$ times the
number of paths of length $m-1$ ending at the oscillators to
which $i$ is coupled. This result is confirmed by the calculation
of the average value of  $J^{(m)}_{ij}$ as an element of ${\cal J
}^m$, under the hypothesis that the elements $J_{ij}$ of $\cal J$
are independent random variables. Assuming, thus, $J^{(m)}_{i1} =
J_0 z^m$, where the constant $J_0$ may sensibly depend on the
specific realization of the random network, we find
\begin{equation} \label{Ai2}
A_i \approx J_0 \sum_{m=d_i}^\infty  (z+{\rm i}\Omega)^{-m-1} z^m
= \frac{J_0}{\Omega} \left(1+\frac{\Omega^2}{z^2} \right)^{-d_i/2}
\exp \left[ -{\rm i} \left(\frac{\pi}{2}+ d_i \tan^{-1}
\frac{\Omega}{z} \right) \right].
\end{equation}
Comparing with eq. (\ref{Ai1}), we realize that the dependence of
$|A_i|$ on $d_i$ is now much slower, especially, for small $\Omega$.
On the other hand, the phase shift between oscillators whose
distances differ by one, $\Delta \varphi = \tan^{-1} (\Omega/z)$, is
the same as for small distances. When $\Omega$ is small, the only
difference is an additional shift of $-\pi/2$. This is clearly
confirmed by numerical results: the dotted lines in fig. \ref{fig1}b
are displaced by $-\pi/2$ with respect to the corresponding full
lines, which stand for the small-$d_i$ analytical prediction. An
independent confirmation of eq. (\ref{Ai2}) is given by the fact
that for small $\Omega$, large $d_i$, and a given connection
network, the amplitude modulus should behave as $|A_i| \sim
\Omega^{-1}$. This is shown in fig. \ref{fig2} for the same network
as in fig. \ref{fig1}, at the maximal distance, $d_i=15$. Note that
this behaviour can be interpreted as a resonance-like phenomenon in
the response of the oscillators --whose natural frequency is
$\omega_i=0$-- to the perturbation. The external harmonic action
induces a larger effect when its frequency is closer to the
individual frequency of the dynamical components of the system.

Though the results discussed so far apply to random oscillator
networks, our analytical approach can be used for arbitrary
connection patterns. In the special case of regular networks we
can even obtain exact results. Consider, for instance, a directed
ring, i.e. a linear array with periodic boundary conditions,
where each oscillator is coupled to its nearest neighbour to the
left ($z=1$). In this situation, $J_{ij}= 1$ if $i = (j+1) \mod N$
and $0$ otherwise, which implies that $J^{(m)}_{i1}$ vanishes
unless $m=d_i+kN$ ($k=0,1,2,\dots$). Explicitly calculating the
amplitude from eq. (\ref{Ai}), we find
\begin{equation}
A_i =\left[ 1-(1+{\rm i}\Omega)^{-N}\right]^{-1} (1+{\rm
i}\Omega)^{-d_i-1}.
\end{equation}
Here, the amplitude modulus decays exponentially for all $d_i$ and
the phase shift between oscillators whose distances differ by one
is constant, $\Delta \varphi=\tan^{-1} \Omega$.

Another situation where the decay of the amplitude modulus is
purely exponential is the case of a tree-like connection network.
Trees are graphs which contain no cycles. In directed trees,
consequently, there is at most one path joining any two nodes,
and therefore $z=1$. Under such conditions, only one term in eq.
(\ref{Ai}) contributes to the amplitude, and
\begin{equation}
A_i =   (1+{\rm i}\Omega)^{-d_i-1} .
\end{equation}
Those nodes which cannot be reached from the perturbed oscillator
have $A_i=0$.

If the number $z_i$ of connections ending at each node is not
uniform over the network,  eq. (\ref{osc3}) can still be reduced
to (\ref{calM}). Now, however, ${\cal M}= {\cal Z}+{\rm i} \Omega
{\cal I}-{\cal J}$ where $\cal Z$ is the diagonal matrix with
elements $Z_{ij}=z_i \delta_{ij}$. In this situation, we cannot
insure that the eigenvalues of the connection matrix $\cal J$ are
bounded in modulus, and therefore we are in general not able to
expand ${\cal M}^{-1}$ in order to obtain an expression similar
to eq. (\ref{Ai}). Whereas we may numerically solve eq.
(\ref{calM}) by inverting $\cal M$, our analytical approach does
not apply anymore. Moreover, for such an arbitrary connection
pattern the stability of the fully synchronized state is not
guaranteed, and has to be separately analyzed for each
realization of the network.

\begin{figure}
\onefigure[width=10cm]{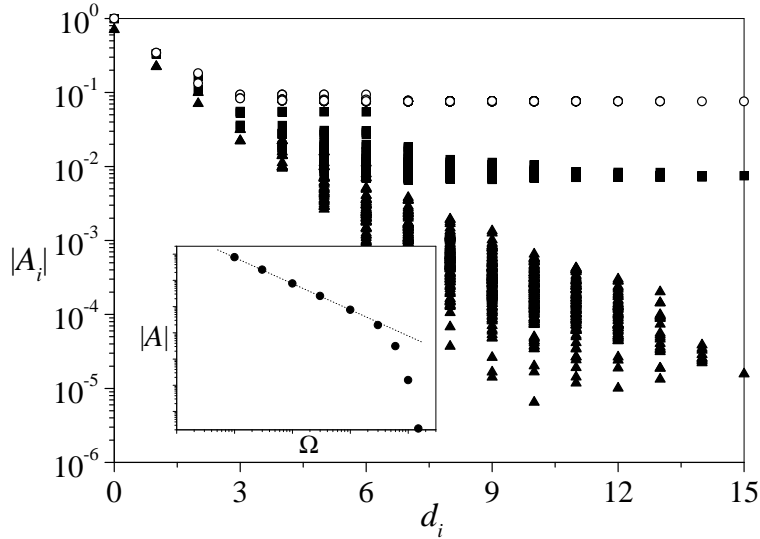} \caption{Amplitude modulus as a
function of the distance from the perturbed oscillator in a
$10^3$-oscillator random network where the number of neighbours
of each node is $z_i=1$, $2$ or $3$ with equal probability.
Symbols are as in fig. \ref{fig1}. The inset shows the amplitude
modulus at the maximal distance, $d_i=15$, as a function of the
perturbation frequency. Axis scales in the inset are the same as
in fig. \ref{fig2}. The dotted line has slope $-1$.} \label{fig3}
\end{figure}

Figure \ref{fig3} shows numerical results for the amplitude modulus
as a function of the distance from the perturbed oscillator, for a
$10^3$-oscillator random network with a random distribution of
$z_i$. Specifically, $z_i$ is chosen to be $1$, $2$, or $3$ with
equal probability, in such a way that the average value of $z_i$
equals the value of $z$ for the network considered in figs.
\ref{fig1} and \ref{fig2}. Once $z_i$ has been determined for each
node, its $z_i$ neighbours are selected at random. For the network
realization of fig. \ref{fig3} the maximal distance is, again,
$d_i=15$. We find that, qualitatively, $|A_i|$ shows the same
dependence on $d_i$ and $\Omega$ as for the regular graph. As may
have been expected, however, there is a larger dispersion in the
amplitude modulus for elements at a given distance from the
perturbed oscillator, especially, at large frequencies. The value of
$|A|$ at the maximal distance, shown in the inset of fig.
\ref{fig3}, exhibits the same dependence on the perturbation
frequency as in the former case.

In summary, we have shown that the response of a network of
synchronized phase oscillators to a local harmonic perturbation of
external origin exhibits two well-defined regimes, depending on the
distance to the point at which the perturbation is applied. At small
distances, the system behaves as a dissipative medium. For a variety
of connection patterns, the perturbation propagates through the
network at constant velocity, while its amplitude decreases
exponentially with the distance. In random networks, this behaviour
holds up to certain threshold distance, beyond which the variation
of the amplitude with the distance is much slower. The smaller the
perturbation frequency, the larger the amplitude at long distances
and the smaller the threshold. Our analytical approach, which
applies to regular graphs, suggest that the existence of these two
regimes are a consequence of the geometrical properties of the
network, associated with the number of paths connecting the
perturbed oscillator  with any other node at a given distance. The
analysis is linear, and the solutions for harmonic perturbations
studied here can be combined to describe any other perturbation of
sufficiently small amplitude. Nonlinear effects in the propagation
of perturbations, which are expected to play a role for stronger
perturbations, will be the subject of a future contribution.

Let us finally point out that the present results establish a
direct link between the dynamics of an ensemble of coupled
oscillators and the geometry of the connection network. This
suggests a method to explore the structure of connections if the
individual activity of oscillators is accessible through
measurements. Determining the response of each oscillator to
external perturbations applied to different elements of the
ensemble could make it possible to reconstruct the underlying
interaction pattern.

\acknowledgments Valuable discussions with G. Abramson, H. Kori,
M. Kuperman, and A. S. Mikhailov and are gratefully acknowledged.

\end{document}